# Cache the Knowledge, Not Just the Prompt: Dynamic Datalog as the Action Language for LLM Agents

Bruno Rucy Carneiro Alves de Lima
rucy@lucidarium.systems
Lucidarium Systems
Tartu, Estonia
University of Tartu
Institute of Computer Science
Tartu, Estonia

Victor Henrique Cabral Pinheiro
victor.pinheiro@ut.ee
University of Tartu
Institute of Computer Science
Tartu, Estonia

## Abstract

An LLM agent that answers questions about a knowledge graph writes a program per question, and its tools evaluate each from scratch. We give the agent a *Datalog REPL*, a session over a dynamic Datalog engine, in which rules are inserted and retracted like facts while derived facts stay current. Prompt caching reuses the model's processing of what a session shares, and the Datalog REPL reuses the evaluation of it. On chains of questions over three knowledge graphs, agents reuse a relation that the REPL keeps in 179 of 180 cases, with two models, and write fewer output tokens: SQL and Python agents write 23 to 82% more. With SQL tables or a Python namespace, Claude Haiku 5.5 stores almost nothing for later questions, and a chain of seven questions at 13 million facts takes it a median of 48 and 56 s, against 31 s with the REPL. Claude Opus 5.5 matches the REPL by storing tables in SQL, and when the data changes it stores views, which are evaluated again at every read. A relation that the REPL keeps is read like a table and stays current like a view. Datalog is a niche language beside SQL and Python, and with thinking enabled the models answer as accurately with it.

## CCS Concepts

• **Computing methodologies → Knowledge representation and reasoning**; *Natural language processing*; • **Information systems → ***Query languages*; *Deductive databases.*

## Keywords

LLM agents, Datalog, Knowledge graphs, Dynamic Datalog, Incremental computation, Tool calling, Prompt caching

## 1 Introduction

LLM agents increasingly act by writing code. In CodeAct [20], the agent emits Python that a read-eval-print loop (REPL) executes, and the REPL keeps variables and definitions across turns, so later code builds on earlier results. Recursive Language Models give the model such a REPL with a long context held as a variable and calls to sub-models offered as functions [23], and programmatic tool calling exposes the tools that allow it as functions that the generated code calls [2].

Over a session, the prompt of such an agent grows. The system prompt stays, and every turn appends the agent's code and the results it produced. Prompt caching exploits this. The model's processing of a prompt, its attention states, can be stored and reused by a later request that contains the same declared segment of text [9]. Commercial APIs store the processing of a prompt's prefix and bill its reuse by a later request at a tenth of the input price or less [3].

The evaluation of the agent's code has no such reuse. Consider an agent that answers questions about a knowledge graph by writing queries. Many questions require recursion ("which organisations belong to this university, directly or indirectly?") or inference from an ontology ("which employees work for this department?", when the data records only full professors and lecturers). Over a session the agent writes a program for every question and revises it when a result looks wrong, and a query engine evaluates every version from scratch over the whole graph, however small the program and however much it shares with the programs before it. Materialized views, which store the results of a program and update them when the data changes, remove this cost for what the ontology implies [11] [15]. The rules a question brings, and every revision of them, are evaluated from scratch by the engines that agents use. An agent can store a derived relation for later questions, as a table or a variable, but the copy is the agent's to manage, and it goes stale when the data changes.

Datalog states such questions compactly [1]. A Datalog program is a set of rules, each of which derives facts from other facts. Recursion is an ordinary rule, an ontology is a set of rules written once, and a question adds a few rules and a query. Over a session, the agent's Datalog program grows the way its prompt grows.

A *dynamic Datalog* engine accepts insertions and retractions of rules, as well as of facts, while it runs, and after each update it reports how the facts that the program derives have changed [6]. Prompt caching reuses the processing of the prompt that a session shares, and a dynamic Datalog engine reuses the evaluation of the program that a session shares. We combine the two. A *Datalog REPL* is a session that holds a knowledge graph, its rules and every rule the agent has written, and applies each step of the agent as an update: the rules of a new question, or a revision of them. The facts that the graph's ontology implies are computed once per session, each step is evaluated as an update to the facts derived before it, and the relations the agent defines persist as named rules, which answer later questions about any entity and stay current when the data changes, with no state for the agent to manage.

We are not aware of prior work that gives an LLM agent a dynamic Datalog engine as its tool, or that relates the evaluation of an agent's programs to prompt caching. Our evaluation has five findings. **Agents use what the session keeps.** With two models, they reuse a relation that the REPL keeps in 179 of the 180 cases where one applies and in 64 of 66 while the data changes, and they write fewer output tokens: the SQL and Python agents write 23 to 82% more. **A small model gets no such reuse from SQL or Python.** With a persistent SQL database or Python namespace, it stores a relation in 2 of 288 cases, and on a graph of 13 million facts its chain of seven questions takes a median of 48 and 56 s, against 31 s with the REPL. **A large model chooses in SQL between a table and a view.** It stores tables when the data is fixed, and its chains then take as long as with the REPL. When the data changes it stores views, which are evaluated again at every read. A relation that the REPL keeps is read like a table and stays current like a view. **The saving grows with the graph.** On all three graphs, a question on top of a kept relation saves more time the larger the graph is, up to 0.63 against 4.69 s. A change to the data costs about 1 ms, and at 13 million facts a one-rule revision of a program that keeps a large relation takes 2 ms as an update against 36 s from scratch. **Datalog is a niche language, and the models are as accurate with it.** With thinking enabled, they answer as accurately in it as in the languages we compare it with, and an open model of 1.7 billion parameters answers recursive questions in Datalog alone.

This paper makes three contributions.

(1) The Datalog REPL is a design for agent tools in which rules and facts are updates, answers are changes to the derived facts, and the derived facts persist across the questions of a session.
(2) A benchmark of 23 questions over three knowledge graphs labels each question by measuring what its reference program needs, and an agent harness poses every question in Datalog, SQL, SPARQL and Python, each over the base facts and over all derived facts.
(3) An evaluation measures whether agents reuse what a session keeps, what that reuse saves against the SQL and Python tools that agents use, what a revision costs, and how accurately LLMs write Datalog.

## 2 Related Work

**Reuse in LLM serving.** Serving systems reuse the model's processing beyond a single prompt prefix. SGLang's RadixAttention caches the attention states of the prefixes that the calls of a language-model program share [24], and semantic caches such as GPTCache store a model's responses by the embedding of the query and return a stored response to a similar query [8]. The Datalog REPL reuses the evaluation of the agent's programs, and a kept rule answers later questions about any entity.

**Agents that keep what they computed.** Voyager stores an ever-growing library of executable code skills that later tasks retrieve [19], and Agent Workflow Memory induces reusable workflows from past tasks [21]. The Datalog REPL keeps the agent's relations as rules whose derived facts the engine keeps current as the data changes.

**LLMs that write logic programs.** Faithful Chain-of-Thought translates a question into a reasoning chain in Python, Datalog or a planning language that a deterministic solver executes, so that the chain is a faithful explanation of the answer [13], and Logic-LM translates a problem into logic programs and other solver inputs and revises them from the solver's error messages [16]. The Datalog REPL keeps the solver's state across a session, so the programs for one question build on those of earlier questions, and each revision is applied as an update.

**Incremental and dynamic Datalog.** DRed and counting [11], and recursive counting and forward/backward/forward maintenance [15], maintain the facts that a program derives while the given facts are inserted and deleted, DRed also while rules are inserted and deleted [11], and IncA maintains the facts that a fixed program analysis derives as the analysed program changes [18]. DDlog compiles a fixed program into an incremental dataflow [17] built on differential dataflow [14], and Soufflé compiles a fixed program into parallel C++ [12]. DBSP derives the incremental version of a query, recursion included, from the query itself [4]. Earlier work evaluates differential dataflow as a backend for Datalog under frequent updates [5] and applies incremental computation to distributed ontology reasoning in which the ontology itself changes [7]. The engine of [6] evaluates dynamic Datalog programs, whose rules change at run time, as a higher-order DBSP program in which the rules themselves are inputs.

**Scope.** The REPL uses an extended implementation of an existing dynamic Datalog engine [6], and our contribution is the design that exposes dynamic Datalog to an agent, together with its measurement. We compare the REPL with the SQL and Python tools that agents use, as the agents used them, on the recursive and derived relations that the questions need, and with no other reasoner.

## 3 Background

### 3.1 Datalog by example

A Datalog program consists of facts and rules. A fact states a relationship between constants, as in `parentof(ann, bob)`. A rule states that a fact follows from other facts. The two rules

```
ancestorof(X, Y) :- parentof(X, Y).
ancestorof(X, Z) :- parentof(X, Y), ancestorof(Y, Z).
```

read: every parent of `Y` is an ancestor of `Y`, and every parent of an ancestor of `Z` is an ancestor of `Z`. Names that start with an uppercase letter are variables. The second rule mentions `ancestorof` on both sides of `:-`, which makes it recursive. Predicates given by facts, such as `parentof`, are *base* predicates, and predicates defined by rules, such as `ancestorof`, are *derived* predicates. An engine evaluates a program by applying its rules to the facts until no new fact follows. The facts obtained this way, beyond the given ones, are the program's *derived facts*, and all facts together are its *closure*. A question is a query over base and derived predicates, possibly with rules of its own. The question "who are the ancestors of cid?" is the query `?- ancestorof(X, cid).`, and the question "which women are ancestors of cid?" becomes

```
answer(X) :- ancestorof(X, cid), female(X).
?- answer(X).
```

The generated family graph of our evaluation starts its series of questions with these two.

A knowledge graph with an ontology takes the same form. In the LUBM university benchmark [10], the data records facts such

```
Question 1 of 5: Who are the ancestors of person115?
ancestorof(X, Y) :- parentof(X, Y).
ancestorof(X, Y) :- parentof(X, Z), ancestorof(Z, Y).
answer(X) :- ancestorof(X, "…/person115").
?- answer(X).
34 rows …
Changes (facts added and removed, by predicate):
- ancestorof/2: +2824 -0 …
- answer/1: +34 -0 …
Library: empty.
Question 2 of 5: Which women are ancestors of person115?
Library: ancestorof/2 (2 rules).
answer(X) :- female(X), ancestorof(X, "…/person115").
?- answer(X).
17 rows …
Changes (facts added and removed, by predicate):
- answer/1: +17 -0 …
Library: ancestorof/2 (2 rules).
```

```
Question 2 of 5, the SQL session:
WITH RECURSIVE anc(a, b) AS (
  SELECT subject, object FROM parentof
  UNION
  SELECT p.subject, anc.b
  FROM parentof p JOIN anc ON p.object = anc.a
)
SELECT DISTINCT anc.a AS entity
FROM anc JOIN female f ON f.entity = anc.a
WHERE anc.b = '…/person115'
```

**Listing 1: Left: two questions of a session whose REPL keeps the agent's predicates, with the agent's programs and the tool's replies. Right: the SQL agent's program for the second question, which states the recursion again. Answer rows, sample facts and the identifiers' prefix are left out.**

as which people work for which departments, and the benchmark's Univ-Bench ontology states, for instance, that every full professor is a professor and that an employee is a person who works for an organisation. Each such statement becomes a rule,

```
professor(X) :- fullProfessor(X).
employee(X) :- person(X), worksFor(X, Y),
organization(Y).
```

and our translation of the LUBM ontology consists of 99 rules, which derive facts that the data never states.

## 3.2 Dynamic Datalog

Keeping the derived facts current while base facts are inserted and deleted is known as incremental view maintenance [11] [15], and compiled engines such as DDlog and Soufflé fix the program when they compile it [17] [12]. A dynamic Datalog engine treats rules as input in the same way as facts. An update inserts or retracts rules and facts, and after each update the engine reports the change in the derived facts: the facts that now follow, and the facts that no longer do. The engine computes this change from the state it keeps between updates, instead of evaluating the program again from scratch [6]. We use `materialized-view`, an extended implementation of the dynamic Datalog engine of [6] written in Rust, through its Python bindings, PyDYRE, at commit `b34efb0` of `github.com/brurucy/materialized-view`. It evaluates the higher-order DBSP program of that work, and we refer to it as the engine.

# 4 A Datalog REPL for agents

## 4.1 Sessions, updates and answers

A session holds the base facts of a knowledge graph, the rules of its ontology, and the rules the agent has written, and a dynamic Datalog engine keeps the derived facts of all of them current. The agent uses one tool, which accepts a fragment of Datalog source and returns the answer to its query together with a summary of what the fragment changed.

**One tool call, step by step.** The REPL handles a call in four steps.

(1) It parses the fragment into rules, facts and one query, and checks them against the graph's schema, the predicates that the session keeps and the engine's limits, at most three terms in an atom and five variables in a rule. A fragment that fails a check is rejected with a message, and the session stays as it was.
(2) It takes the set difference with the program that it recorded for the current question, and applies it to the engine as one update, which retracts the rules and facts that the fragment drops and inserts those that it adds.
(3) It reads the answer to the query from the derived facts.
(4) It returns the answer rows and a summary of the update: for each predicate whose facts changed, the number of facts added and removed, with up to three examples, for at most ten predicates, the largest change first.

Listing 1 shows two calls of one agent, on the two questions of the previous section, asked about person115 of a generated family graph.

The state of a Python REPL is the namespace that results from executing statements in a particular order, and the state of a Datalog REPL is the set of derived facts of the current rules and facts. A revision is a difference. If the agent first omits `female(X)` from the `answer` rule and adds it after seeing the result, the second step retracts the old `answer` rule and inserts the new one, and the `ancestorof` facts, which both versions share, are not derived again. The agent therefore writes complete programs, as it would in SQL or SPARQL, and the engine processes only their difference. A rule ends with a period and is undone exactly by its retraction, so the REPL can also apply each rule as soon as the model has generated it.

**The end of a question.** Questions share a session in one of three modes, and in all of them the knowledge graph and its rules are loaded once. In the *isolated* mode, the REPL retracts a question's rules when the question ends, so every question sees only the graph and its own rules. In the *library* mode, the agent marks the predicates that later questions may use with a `.share` directive in its program. When the question ends, the REPL keeps the rules of the shared predicates and of the predicates they read, and retracts the question’s other rules. In the *keep-all* mode, the REPL keeps the rules of every predicate except `answer` and the predicates that read it, with no marking. Listing 1 is a keep-all session: the first question leaves the two `ancestorof` rules in the session, and the program of the second question is one rule that reads them. Every result and every new question lists the kept predicates. The REPL counts the holders of each rule, so a rule that a question states again is inserted once, and it leaves the engine when its last holder releases it.

**Table 1: What changes during an agent session over a knowledge graph, and which kind of system processes each change as an update. “No” means the system evaluates again from scratch.**

| | Graph facts | A question’s rules | A revision | Kept for the next question |
|---|---|---|---|---|
| Query engine, each query from scratch | no | no | no | nothing |
| Materialized views, maintained incrementally [15] [17] | yes | no | no | what the ontology implies |
| Python REPL (CodeAct, Recursive Language Models) | no | no | no | values the code stored |
| **Datalog REPL** | **yes** | **yes** | **yes** | **all derived facts** |

**Table 2: Knowledge graphs, and their questions by what each question’s own program adds to the graph’s predicates.**

| Graph | Base facts | Facts with rules | Rules | Questions | Lookup | Join | Recursion |
|---|---|---|---|---|---|---|---|
| LUBM-3 | 336,130 | 462,719 | 99 | 12 | 3 | 8 | 1 |
| Follows | 921 | 21,458 | 3 | 4 | 1 | 3 | 0 |
| Family | 453 | 8,567 | 9 | 7 | 2 | 3 | 2 |

## 4.2 How an update is evaluated

The engine has two inputs, the facts and the rules, and each input is a set of changes in which an insertion has the weight +1 and a retraction the weight −1 [6]. A rule enters as data. One entry per body atom records which atom a *partial match* of the rule, a binding of the variables matched so far, meets next, and which columns of that atom are bound by then, by a constant or by an earlier atom. One entry records the head to derive once the body is matched. Evaluation is a loop of four joins. Every fact is indexed by the bound columns of the body atoms over its predicate, a partial match is paired with the next atom of its rule, the indexed facts that agree with it on the bound columns extend it, and a match that has passed the last body atom yields its head as a new fact, until no new fact appears. Each join is a DBSP operator with incremental semantics, which computes the change of its output from the changes of its inputs and the state it keeps. An update therefore enters the loop as a change, and the engine reports the change in the derived facts, which the REPL’s summary counts by predicate. In Listing 1, the rule of the second question enters as one change and leaves as 17 new `answer` facts, and the summary lists no change to `ancestorof`.

The REPL’s dialect is positive Datalog, the fragment that [6] evaluates.

## 4.3 What a session changes and what it caches

Over a session, four things change: the graph’s facts, the rules of each new question, the revisions of those rules, and the work available to the next question. Table 1 lists which of them each kind of system that agents use today processes as an update.

Prompt caching and the Datalog REPL combine in one session: the prompt that describes the graph’s schema and rules is cached on the model side, and the graph’s derived facts are cached on the evaluation side. A prompt cache reuses the model’s processing of a prefix that an earlier request shares, and everything after the first changed token is processed again. The REPL reuses the derived facts of every rule already in the session, in any order, and a change updates only the consequences of the changed rules.

# 5 Evaluation

We ask four questions. RQ1: do agents reuse what a session keeps, across questions and while the data changes? RQ2: what does the reuse save, against the SQL and Python tools that agents use? RQ3: what does a revision cost? RQ4: are LLMs as accurate when they write Datalog, a niche language, as when they write SQL, SPARQL and Python?

## 5.1 Knowledge graphs and questions

We use three knowledge graphs (Table 2). LUBM-3 consists of universities 0 to 2 of the LUBM benchmark [10] with the Univ-Bench ontology. Follows is a generated network of 12 communities of 25 accounts each, with cycles inside every community and one-way links between communities. Family is a generated genealogy of 159 people over six generations, the graph of Listing 1. Together, the questions of the three graphs require joins, ontology inference and recursion.

Every question has a reference program in Datalog, and every label of a question is a measurement on its reference program. What a question adds to the graph’s predicates is a *lookup* when its program has no rules of its own, a *join* when it has rules without recursion, and *recursion* when one of its rules is recursive. The build of the question set rejects a question if removing any one condition of its own rules, or its own recursive rules, leaves the answer unchanged. Every gold answer is computed twice, by the Datalog reference program and by a SPARQL reference query over the graph with all derived facts added, and the two must agree. The derived facts of LUBM-3 match those of an independent OWL 2 RL reasoner for all 41 classes and properties, and those of the generated graphs match plain Python implementations of every derived predicate.

## 5.2 Agent harness and conditions

Eight conditions combine one of four languages with one of two kinds of data. In the conditions marked M, programs run over the graph with all derived facts available, which is what a Datalog REPL holds once the graph's rules are loaded, so the agent writes only what the question adds: a query and helper rules in Datalog, SQL, SPARQL 1.1, or a Python function that returns the answer rows. In the conditions marked B, programs run over the base facts alone, and the agent also writes the graph's reasoning. Datalog runs in a Datalog REPL, SQL in DuckDB 2.0.0, SPARQL in Oxigraph (pyoxigraph 0.5.13), and Python in a sandbox without network access.

Every condition receives a system prompt with the same sections in the same order, and the graph's rules are stated as the same English sentences in every condition, so that no language receives them in its own syntax. In an agent loop that interleaves reasoning and tool calls [22], the model writes a program, runs it at most three times through the condition's tool, reads each result, and submits a final program. In the Datalog conditions, the REPL applies each run as an update to the program of the previous run, and the agent is not told so. The scorer runs the submitted program again and compares its answer rows with the gold answers as sets. We report the rate of exact matches and the F1 score over answer rows. The model is Claude Haiku 5.5 with adaptive thinking at high effort, each question is run three times, each run an *epoch*, and the harness is inspect-ai 0.3.285. The independent questions are also run in two further settings: the same model with thinking turned off, and LFM3-1.7B, an open model of 1.7 billion parameters, run locally in one epoch.

## 5.3 Sessions

Independent samples measure accuracy, and they leave nothing for a later question to reuse. A session poses several questions about one graph to one agent, one after the other, in one conversation and with one REPL, in the Datalog-B condition, so the agent writes every rule that a question needs, including the graph's rules.

The independent questions share few rules beyond the graph's, so we also pose *chains* of questions about one graph that build on one another: seven on LUBM-3, four on Follows and five on Family, whose first two questions are those of Listing 1. Each question is self-contained, and 12 of the 16 have a reference program over the base facts that defines a predicate with exactly the rules of an earlier question, such as the reachability relation of Follows, asked about account30 and later about account130. A question reuses an earlier predicate *by name* when its program reads a kept predicate without defining it, and *by its rules* when it states a kept predicate with the same rules, an empty update. Otherwise it *restates* an earlier question's rules, which the engine derives again, or defines what it needs anew.

We compare the Datalog sessions with two sessions that keep what the agent computes: a Python REPL whose namespace persists across questions, as in CodeAct [20], and a DuckDB database whose tables and views persist. A view is a named query that the database evaluates again at every read, and DuckDB has no view that it keeps up to date. In the chains, their prompts state, as the Datalog prompts do, that a relation computed for one question is often needed again, and their results list what earlier questions stored. Each chain is posed twelve times in all five kinds of session with Claude Haiku 5.5, and three times in the isolated, keep-all, SQL and Python sessions with Claude Opus 5.5, a larger model of the same family, with the same thinking setting. A question in one session is a *case*, so the 12 questions that can reuse give 144 cases for each kind of session with Haiku and 36 with Opus.

## 5.4 Measurement and statistics

**Replays.** To time the engine on larger graphs, we replay the agents' LUBM-3 programs, in the order each sample ran them, on LUBM-10, LUBM-30 and LUBM-100, universities 0 to 9, 0 to 29 and 0 to 99 of the same generator, with the same 99 rules and 1.3, 4.0 and 13.4 million base facts. Scored against gold answers computed at each size, the co-authorship question (authors connected through chains of co-authorships) aside at LUBM-100, every Datalog program that is correct on LUBM-3 is correct there, and every wrong one stays wrong. Every time we report for LUBM is this replay on LUBM-100, with eight worker threads of the engine, on an Apple M5 Max with 18 cores and 48 GB of memory. The sessions on Follows and Family are replayed on their own graphs and on larger graphs from the same generators: Follows with 10, 30 and 100 times the communities, and Family as 100 and 1,000 genealogies. A time is a single run unless we give the number of runs. A question's *engine time* in a session is the time of its runs and of the update that ends the question. Times that depend on the model are measured by sending its requests again through the streaming API, with the tool's input streamed as it is generated and the thinking text omitted. The calls of all kinds of session of a model are sent in the same minutes, one kind beside the other. In every replay of this paper, the update and the evaluation from scratch give the same answer.

**Statistics.** Comparisons are paired by the unit named with them, and tests are two-sided: an exact sign test or a Wilcoxon signed-rank test for paired measurements, an exact Mann-Whitney test for two groups of sessions, an exact McNemar test and Cochran's Q for paired correctness, Fisher's exact test for counts, and Holm's correction within a family. Intervals are 95% bootstrap intervals.

## 5.5 RQ1: agents reuse what the session keeps

**Reuse.** With Claude Haiku 5.5 (Table 3), every case of the keep-all sessions that can reuse an earlier predicate does so, 98 by name and 46 by its rules, and in the library sessions 99 of the 144 do so, through the predicates the agent shared. The Python and SQL sessions ran and submitted 786 programs. No SQL query stores a

**Table 3: Chain sessions with two models. With Claude Haiku 5.5, the keep-all agents reuse an earlier predicate in every case and on every graph, and the Python and SQL agents in 2 of 288 cases. With Claude Opus 5.5, the keep-all and the SQL agents reuse in 35 of 36 cases. A Python or SQL question restates when three of its code lines repeat an earlier question's.**

| | | Of the cases that can reuse | | | Cases that reuse, per graph | | | | |
|---|---|---|---|---|---|---|---|---|---|
| Session | Exact | By name | By its rules | Restated | Family | Follows | LUBM-3 | Rules | Output tokens |
| Claude Haiku 5.5, twelve sessions of each kind per chain, 144 cases that can reuse | | | | | | | | | |
| Isolated | 192/192 | 0 | 0 | 144 | 0/48 | 0/36 | 0/60 | 5.0 | 705 |
| Library | 191/192 | 96 | 3 | 44 | 31/48 | 27/36 | 41/60 | 3.2 | 669 |
| Keep-all | 192/192 | 98 | 46 | 0 | 48/48 | 36/36 | 60/60 | 2.8 | 582 |
| Python | 192/192 | 2 | | 140 | 2/48 | 0/36 | 0/60 | | 849 |
| SQL | 192/192 | 0 | | 114 | 0/48 | 0/36 | 0/60 | | 919 |
| Claude Opus 5.5, three sessions of each kind per chain, 36 cases that can reuse | | | | | | | | | |
| Isolated | 48/48 | 0 | 0 | 35 | 0/12 | 0/9 | 0/15 | 5.4 | 403 |
| Keep-all | 48/48 | 35 | 0 | 0 | 12/12 | 9/9 | 14/15 | 3.0 | 333 |
| Python | 48/48 | 12 | | 24 | 0/12 | 3/9 | 9/15 | | 607 |
| SQL | 48/48 | 35 | | 1 | 12/12 | 9/9 | 14/15 | | 412 |

table for a later question, and one Python session defines a function that two later questions call. These sessions restate the earlier code in 140 and 114 of the 144 cases. Paired by question, the keep-all agents reuse more often than the isolated, library, Python and SQL agents in all 12 questions that can reuse (sign test, $p < 0.001$ each). No session defines a kept predicate with other rules.

**A larger model.** With Claude Opus 5.5, the keep-all agents reuse an earlier predicate in 35 of the 36 cases, and so do the SQL agents. In each of their nine sessions, the SQL agents store a relation, such as the ancestors of every person, as a table or as a view, and their later queries read it. The Python agents read a stored name in 12 cases, in four of their nine sessions.

**Shorter programs.** With the predicates kept, a Haiku question needs fewer rules than in the isolated sessions and fewer output tokens in all 12 questions that can reuse (sign test, $p < 0.001$ each). Against the Python and SQL sessions, the keep-all agents write fewer output tokens in 15 of the 16 questions (Wilcoxon signed-rank test, $p = 0.001$ and $p < 0.001$): 582 per question against 849 and 919, a mean of 267 fewer than Python (95% interval 148 to 386) and 337 fewer than SQL (206 to 480). The Opus keep-all agents write 333 output tokens per question against 607 in Python and 412 in SQL, fewer in 16 and 15 of the 16 questions ($p < 0.001$ each), where the SQL agents reuse as often.

**Changing data.** With Haiku, we also pose the chains three times with base facts inserted and retracted between questions, eight changes in all, after which 11 of the 16 questions would be answered wrongly from a stored copy of what they reuse. Each change is applied through the session's own tool, and the next question lists the facts added and removed. Agents whose REPL kept their predicates reused one in all 33 cases of these questions, more often than the isolated, Python and SQL agents in all 11 questions (sign test, $p = 0.001$), and all 240 answers over the five kinds of session were exact. On LUBM-100, a change costs the keep-all session a median of 0.9 ms against 31 ms to derive its kept predicates again, less in all 9 changes ($p = 0.004$), and the question after it 25 against 197 ms.

With Opus, in three sessions of each of four kinds, all 192 answers were exact, and the keep-all agents reused a kept predicate in 31 of the 33 cases. The SQL agents, whose prompt states that the data can change, defined views and no table in all nine sessions, and read a view in 30 of the 33 cases.

## 5.6 RQ2: what the reuse saves

**The agents' chains.** Table 4 sets the keep-all sessions beside the SQL and Python sessions on LUBM-100. With Haiku, the two questions on top of the co-authorship relation take the keep-all sessions a median of 48 and 25 ms of engine time. The isolated sessions take 722 and 755 ms, of which 335 and 350 ms derive the relation again and the rest retracts it when the question ends, and the library sessions take 416 and 452 ms. The SQL sessions take 3.8 and 3.3 s and the Python sessions 5.2 and 5.4 s, and both compute the relation again in every question. One SQL session starts its recursion from the one author of the question, and its three co-authorship questions take 51 to 58 ms. In two of the twelve keep-all sessions the agent derived the co-authorship chains of every author, where the other ten derived those of one author. In the first of the two, this closure takes 18.4 s to insert, and the two later questions then take 0.18 and 0.24 s against 31.5 and 30.3 s with the closure derived again, medians of three runs.

With Opus, two of the three SQL sessions store the co-authorship pairs as a table, and their two later questions take 41 to 95 ms, against 4.0 to 4.6 s in the third. The keep-all sessions take 11 to 34 ms, and the Python sessions a median of 7.7 and 10.6 s. When the data changes, no SQL session of Opus stores a table, and each computes the co-authorship relation again in every question: the three co-authorship questions take them 3.9, 4.5 and 3.9 s, against 106, 69 and 38 ms in the keep-all sessions, where a change costs 0.9 ms. A predicate that the REPL keeps is thus read like a table and stays current like a view.

**End to end.** With each agent's generation time, measured by sending its model calls again, the LUBM chain of seven questions takes the Haiku keep-all sessions a median of 31.2 s: 25.4 to 36.7 s in ten sessions, and 45.0 and 51.2 s in the two sessions that derived the closure over all authors. The SQL sessions take a median of 48.2 s and the Python sessions 56.3 s (Table 4). Over the twelve sessions of each kind, the keep-all chains are a mean of 15.6 s shorter than the SQL chains (95% interval 10.3 to 20.1 s) and 22.3 s shorter than the Python chains (17.8 to 26.3 s), and the difference is significant for both (exact Mann-Whitney test, $p < 0.001$ each). Per question, the keep-all sessions take less time than the SQL sessions and than the Python sessions in all 7 questions (sign test, $p = 0.016$ each). With Opus, the keep-all and the SQL chains take a median of 46.0 and 45.2 s, and the Python chains 96.8 s.

**Table 4: The Datalog REPL in keep-all mode against the SQL and Python sessions on the chains, each agent with its own programs. Times are on LUBM-100, as the median over the sessions with the range: a question's engine or execution time, and the chain with the agent's generation. With Haiku, the Datalog agents answer the recursive questions and the chain in less time. With Opus, the SQL agents store tables, and their chains take as long as the Datalog agents'.**

| Measure | Datalog REPL | SQL session | Python session |
|---|---|---|---|
| Claude Haiku 5.5, twelve sessions per tool | | | |
| The co-authorship question | 0.34 s (0.31 to 18.6) | 3.30 s (0.05 to 4.07) | 6.75 s (5.19 to 7.29) |
| The first question on top of it | 48 ms (22 to 324) | 3.83 s (0.05 to 5.01) | 5.24 s (4.98 to 5.65) |
| The second question on top of it | 25 ms (13 to 237) | 3.29 s (0.06 to 4.18) | 5.37 s (4.95 to 5.87) |
| The chain of seven questions | 31.2 s (25.4 to 51.2) | 48.2 s (37.5 to 57.3) | 56.3 s (51.1 to 63.2) |
| Claude Opus 5.5, three sessions per tool | | | |
| The co-authorship question | 0.37 s (0.13 to 0.38) | 0.18 s (0.16 to 3.98) | 9.20 s (8.12 to 9.72) |
| The first question on top of it | 32 ms (31 to 34) | 50 ms (41 to 4,647) | 7.74 s (7.57 to 7.93) |
| The second question on top of it | 27 ms (11 to 30) | 95 ms (53 to 4,038) | 10.6 s (8.59 to 11.6) |
| The chain of seven questions | 46.0 s (45.9 to 47.8) | 45.2 s (44.9 to 58.9) | 96.8 s (83.3 to 99.9) |

**Where the gain comes from.** With Haiku, the 17 s between the SQL and the keep-all chains are about 8 s of generation, since the keep-all agents write fewer tokens, and 9 s of execution. The chain takes the isolated sessions a median of 36.7 s, more than the keep-all sessions (exact Mann-Whitney test, $p = 0.02$), and the step between the two is the reuse: the keep-all sessions take a mean of 0.7 s less per question (95% interval 0.1 to 1.5 s, Wilcoxon signed-rank test, $p = 0.047$). Against its own last program evaluated once from scratch, a keep-all question takes a median of 3.93 s against 4.29 s, and more than 5% less in 26 of the 84 questions, all of them co-authorship questions: by 6 to 14% in ten sessions, and by 32 to 90% in the two sessions that kept the closure.

**Graph size.** Replayed at the four sizes of LUBM, the runs of a Haiku keep-all question cost a median of 5.0, 5.7, 8.3 and 22 ms on top of the kept predicates against 12, 25, 61 and 195 ms with its predicates from scratch, less in 70 of the 84 questions at LUBM-100 (Wilcoxon signed-rank test, $p < 0.001$). On Follows with 10, 30 and 100 times the communities, 9, 28 and 92 thousand facts whose reachability relation has 0.7, 3.7 and 18.3 million pairs, a question after the first takes the Haiku keep-all sessions a median of 55 ms, 205 ms and 0.63 s of engine time against 191 ms, 908 ms and 4.69 s from scratch. On Family as 100 and 1,000 genealogies, 47 and 447 thousand facts, it takes 22 and 160 ms against 47 and 396 ms. The session takes less time in every case at each of these sizes: all 36 on Follows, 9 of 9 at 100 times, where three sessions are replayed, and all 48 on Family (sign test, $p \leq 0.004$). On the two graphs at their own size, it takes 14 against 12 ms and 7 against 4 ms.

## 5.7 RQ3: cost of a revision

**The agents' revisions.** A revision is a program that an agent runs after an earlier one for the same question. The agents of the independent questions ran 61 revisions in the three settings: 33 on LUBM, 31 of which we replay on LUBM-100, and 28 on Family and Follows. We apply each as the harness does, as one update to an engine that holds the agent's previous program, and compare it with the revised program evaluated from scratch over the persisted data of its condition. On LUBM-100, the 23 revisions that drop rules take a median of 73 ms as updates against 37 ms from scratch, because the update retracts the facts of the dropped rules, which the engine from scratch never derived (Wilcoxon signed-rank test, $p = 0.11$). The 7 that resubmit the previous program unchanged are empty updates, 1.2 ms against 41 ms ($p = 0.03$). On Family and Follows, the 28 revisions take a median of 3.4 ms as updates against 1.4 ms from scratch ($p = 0.002$).

**One changed rule.** A *single-rule revision* changes one rule of a program and keeps the others: the rule that defines the query's predicate loses its last condition, or, when it has one condition, a constant of the program names the neighbouring entity. We apply it to every LUBM program of the agents at the four graph sizes, as an update to the engine that holds the program and as the revised program from scratch over the persisted data. Figure 1 gives the medians over 155 revisions per size (151 at LUBM-100), beside those of 143 first evaluations (137 at LUBM-100), a sample's first accepted program inserted as an update.

A first evaluation costs the same in both forms: at LUBM-100 the median ratio of the time from scratch to the time of the update is 1.01 (95% interval 0.99 to 1.03). A single-rule revision costs the changed rule as an update and the whole program from scratch, so the gap grows with what the kept rules derive. The update is faster in 110, 131, 135 and 132 of the revisions at the four sizes, by a median factor of 1.1, 1.4, 1.8 and 2.0 (Page's trend test over the 151 revisions measured at every size, $p < 0.001$). At LUBM-100 the factor has a 95% interval of 1.7 to 4.0, and by question it runs from 1.0 to 154. The general co-authorship program, whose kept rules derive the co-authorship chains of every author, is the extreme: its revision takes 1.9, 2.2, 1.6 and 2.1 ms as an update against 0.33,

**Table 5: Accuracy in three settings: Claude Haiku 5.5 with adaptive thinking and with thinking turned off (69 samples per condition) and LFM3-1.7B (23 samples). Exact matches, the test across the four languages on each kind of data, and the mean output tokens per sample, thinking included.**

| | Exact | | | Output tokens per sample | | |
|---|---|---|---|---|---|---|
| **Condition** | **Thinking** | **No thinking** | **LFM3-1.7B** | **Thinking** | **No thinking** | **LFM3-1.7B** |
| Datalog-M | 66/69 | 62/69 | 12/23 | 569 | 324 | 12,352 |
| SQL-M | 69/69 | 69/69 | 12/23 | 649 | 365 | 9,215 |
| SPARQL-M | 67/69 | 66/69 | 9/23 | 710 | 327 | 8,515 |
| Python-M | 68/69 | 69/69 | 9/23 | 700 | 469 | 8,147 |
| Datalog-B | 65/69 | 60/69 | 9/23 | 948 | 466 | 16,498 |
| SQL-B | 66/69 | 63/69 | 12/23 | 916 | 503 | 10,716 |
| SPARQL-B | 67/69 | 55/69 | 8/23 | 1,084 | 469 | 8,887 |
| Python-B | 67/69 | 67/69 | 9/23 | 923 | 577 | 8,534 |
| Cochran's Q, M: $p$ | 0.29 | 0.002 | 0.48 | | | |
| Cochran's Q, B: $p$ | 0.72 | 0.003 | 0.52 | | | |

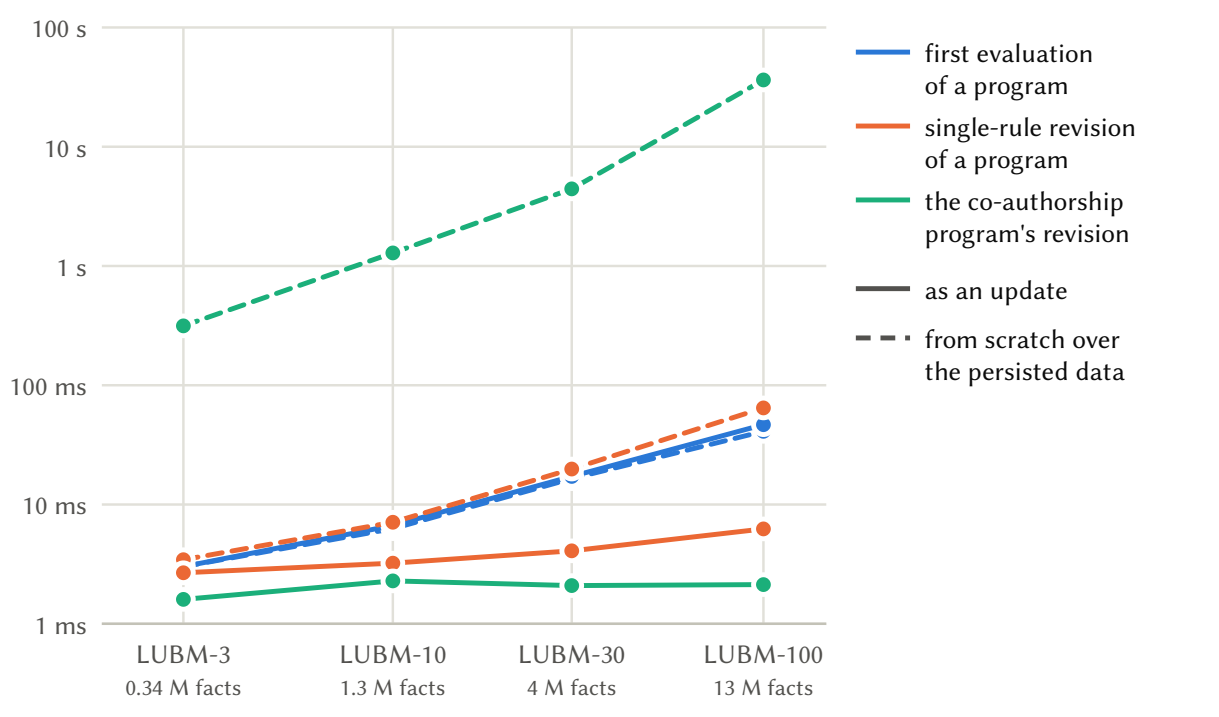

**Figure 1: The agents' LUBM programs replayed at four graph sizes, medians per program, each as an update (solid) and as the same program from scratch over the persisted data (dashed).**

1.39, 6.27 and 36.3 s from scratch, medians of five runs and of three at LUBM-100.

### 5.8 RQ4: accuracy in a niche language

We expect Datalog to be far rarer in a model's training data than SQL or Python, and with thinking the models are nevertheless as accurate with it. In the chains of both models, the keep-all, SQL and Python agents answer every question exactly (Table 3). On the independent questions (Table 5), every condition answers 65 to 69 samples exactly with thinking, and the four languages do not differ on either kind of data (Cochran's Q, $p = 0.29$ and 0.72). Compared with each other language on the same data by an exact McNemar test over the 69 pairs of a question and an epoch, neither Datalog condition differs significantly ($p \geq 0.25$ in the six comparisons): the Datalog conditions answer 1.4 to 4.3 points fewer samples, with 95% intervals between −11.6 and +4.3 points. Without thinking, every condition answers 55 to 69 samples exactly, and the four languages differ ($p = 0.002$ and 0.003). After Holm's correction, no difference between a Datalog condition and another language is significant ($p \geq 0.09$). LFM3-1.7B answers 8 to 12 of the 23 samples of each condition exactly, and the four languages do not differ ($p =$ 0.48 and 0.52). Only in Datalog does it answer any of the three recursion questions: 3 of its 6 attempts, against none of the 18 in the other languages (Fisher's exact test, $p = 0.01$).

### 5.9 Generality

The reuse of RQ1 holds on each of the three graphs, with the data fixed and with the data changing, and with two models (Table 3). The time that a kept predicate saves grows with the graph on all three graphs, and with what the kept rules derive (Figure 1). The single-rule revision is faster as an update at LUBM-100 in each of the three model settings, by a median factor of 2.6, 1.9 and 2.9, and on both kinds of data, by 2.0 and 2.1.

## 6 Conclusion

A Datalog REPL gives an LLM agent a tool whose state is the set of facts that follow from the rules it has written, kept current by a dynamic Datalog engine. Prompt caching reuses the model's processing of what a session shares, and the Datalog REPL reuses the evaluation of it. Agents reuse a predicate that the REPL keeps in 179 of 180 cases with two models, and in 64 of 66 while the data changes, and the SQL and Python agents write 23 to 82% more output tokens. With SQL tables or a Python namespace, a small model stores almost nothing for later questions, and at 13 million base facts its chain of seven questions takes a median of 48 and 56 s against 31 s with the REPL. A large model matches that time by storing tables in SQL, and when the data changes it stores views, which are evaluated again at every read. A predicate that the REPL keeps is read like a table and stays current like a view, and the time it saves grows with the graph on all three graphs. Datalog is a niche language beside SQL and Python, and with thinking the models answer as accurately with it.

## Acknowledgments

This work was supported by the Estonian Research Council grant (PRG2764).

## References

[1] Serge Abiteboul, Richard Hull, and Victor Vianu. 1995. *Foundations of Databases*. Addison-Wesley.

[2] Anthropic. 2026. Programmatic tool calling. Claude API documentation, https://platform.claude.com/docs/en/agents-and-tools/tool-use/programmatic-tool-calling.

[3] Anthropic. 2026. Prompt caching. Claude API documentation, https://platform.claude.com/docs/en/build-with-claude/prompt-caching.

[4] Mihai Budiu, Tej Chajed, Frank McSherry, Leonid Ryzhyk, and Val Tannen. 2023. DBSP: Automatic Incremental View Maintenance for Rich Query Languages. *Proceedings of the VLDB Endowment* 16, 7 (2023), 1601–1614. doi:10.14778/3587136.3587137

[5] Bruno Rucy Carneiro Alves de Lima, Merlin Kramer, and Kalmer Apinis. 2023. On The Suitability of Differential Dataflow For Datalog Interpretation In Highly Dynamic Settings. In *Proceedings of the 2023 6th Artificial Intelligence and Cloud Computing Conference (AICCC 2023)*. ACM, Kyoto, Japan, 218–225. doi:10.1145/3639592.3639622

[6] Bruno Rucy Carneiro Alves de Lima, Merlin Kramer, Kalmer Apinis, and Kristopher Micinski. 2024. Incremental Evaluation of Dynamic Datalog Programs as a Higher-order DBSP Program. In *Proceedings of the 5th International Workshop on the Resurgence of Datalog in Academia and Industry (Datalog-2.0 2024) (CEUR Workshop Proceedings, Vol. 3801)*. 2–16. https://ceur-ws.org/Vol-3801/paper1.pdf

[7] Bruno Rucy Carneiro Alves de Lima, Merlin Kramer, and Victor Henrique Cabral Pinheiro. 2024. Distributed Incremental Ontology Reasoning over Dynamic T-boxes. In *Proceedings of the 2024 7th Artificial Intelligence and Cloud Computing Conference (AICCC 2024)*. ACM, Tokyo, Japan, 423–428. doi:10.1145/3719384.3719446

[8] Bang Fu and Di Feng. 2023. GPTCache: An Open-Source Semantic Cache for LLM Applications Enabling Faster Answers and Cost Savings. In *Proceedings of the 3rd Workshop for Natural Language Processing Open Source Software (NLP-OSS 2023)*. 212–218. doi:10.18653/v1/2023.nlposs-1.24

[9] In Gim, Guojun Chen, Seung-seob Lee, Nikhil Sarda, Anurag Khandelwal, and Lin Zhong. 2024. Prompt Cache: Modular Attention Reuse for Low-Latency Inference. In *Proceedings of Machine Learning and Systems (MLSys)*. arXiv:2311.04934

[10] Yuanbo Guo, Zhengxiang Pan, and Jeff Heflin. 2005. LUBM: A Benchmark for OWL Knowledge Base Systems. *Journal of Web Semantics* 3, 2-3 (2005), 158–182. doi:10.1016/j.websem.2005.06.005

[11] Ashish Gupta, Inderpal Singh Mumick, and V. S. Subrahmanian. 1993. Maintaining Views Incrementally. In *Proc. ACM SIGMOD International Conference on Management of Data*. 157–166. doi:10.1145/170035.170066

[12] Herbert Jordan, Bernhard Scholz, and Pavle Subotić. 2016. Soufflé: On Synthesis of Program Analyzers. In *Proc. International Conference on Computer Aided Verification (CAV) (LNCS, Vol. 9780)*. 422–430. doi:10.1007/978-3-319-41540-6_23

[13] Qing Lyu, Shreya Havaldar, Adam Stein, Li Zhang, Delip Rao, Eric Wong, Marianna Apidianaki, and Chris Callison-Burch. 2023. Faithful Chain-of-Thought Reasoning. In *Proceedings of the 13th International Joint Conference on Natural Language Processing and the 3rd Conference of the Asia-Pacific Chapter of the Association for Computational Linguistics (Volume 1: Long Papers)*. Association for Computational Linguistics, Nusa Dua, Bali, 305–329. doi:10.18653/v1/2023.ijcnlp-main.20

[14] Frank McSherry, Derek G. Murray, Rebecca Isaacs, and Michael Isard. 2013. Differential Dataflow. In *Proceedings of the 6th Biennial Conference on Innovative Data Systems Research (CIDR)*.

[15] Boris Motik, Yavor Nenov, Robert Piro, and Ian Horrocks. 2019. Maintenance of Datalog Materialisations Revisited. *Artificial Intelligence* 269 (2019), 76–136. doi:10.1016/j.artint.2018.12.004

[16] Liangming Pan, Alon Albalak, Xinyi Wang, and William Yang Wang. 2023. Logic-LM: Empowering Large Language Models with Symbolic Solvers for Faithful Logical Reasoning. In *Findings of the Association for Computational Linguistics: EMNLP 2023*. Association for Computational Linguistics, Singapore, 3806–3824. doi:10.18653/v1/2023.findings-emnlp.248

[17] Leonid Ryzhyk and Mihai Budiu. 2019. Differential Datalog. In *Proc. Datalog 2.0 Workshop (CEUR Workshop Proceedings, Vol. 2368)*. 56–67.

[18] Tamás Szabó, Gábor Bergmann, Sebastian Erdweg, and Markus Voelter. 2018. Incrementalizing Lattice-Based Program Analyses in Datalog. *Proceedings of the ACM on Programming Languages* 2, OOPSLA (2018), 139:1–139:29. doi:10.1145/3276509

[19] Guanzhi Wang, Yuqi Xie, Yunfan Jiang, Ajay Mandlekar, Chaowei Xiao, Yuke Zhu, Linxi Fan, and Anima Anandkumar. 2024. Voyager: An Open-Ended Embodied Agent with Large Language Models. *Transactions on Machine Learning Research* (2024). arXiv:2305.16291

[20] Xingyao Wang, Yangyi Chen, Lifan Yuan, Yizhe Zhang, Yunzhu Li, Hao Peng, and Heng Ji. 2024. Executable Code Actions Elicit Better LLM Agents. In *Proceedings of the 41st International Conference on Machine Learning (ICML)*. arXiv:2402.01030

[21] Zora Zhiruo Wang, Jiayuan Mao, Daniel Fried, and Graham Neubig. 2025. Agent Workflow Memory. In *Proceedings of the 42nd International Conference on Machine Learning (ICML) (PMLR, Vol. 267)*. 63897–63911.

[22] Shunyu Yao, Jeffrey Zhao, Dian Yu, Nan Du, Izhak Shafran, Karthik Narasimhan, and Yuan Cao. 2023. ReAct: Synergizing Reasoning and Acting in Language Models. In *International Conference on Learning Representations (ICLR)*. arXiv:2210.03629

[23] Alex L. Zhang, Tim Kraska, and Omar Khattab. 2025. Recursive Language Models. arXiv:2512.24601 [cs.AI]

[24] Lianmin Zheng, Liangsheng Yin, Zhiqiang Xie, Chuyue Sun, Jeff Huang, Cody Hao Yu, Shiyi Cao, Christos Kozyrakis, Ion Stoica, Joseph E. Gonzalez, Clark Barrett, and Ying Sheng. 2024. SGLang: Efficient Execution of Structured Language Model Programs. In *Advances in Neural Information Processing Systems 37 (NeurIPS 2024)*. arXiv:2312.07104